# Probability Based Adaptive Invoked Clustering Algorithm in MANETs

S.Rohini and K.Indumathi
Department of Information Technology, Thiagarajar College Of Engineering
rohinitceit@gmail.com & mathiit22@gmail.com

## ABSTRACT

*A mobile ad hoc network (MANET), is a self-configuring network of mobile devices connected by wireless links. In order to achieve stable clusters, the cluster-heads maintaining the cluster should be stable with minimum overhead of cluster re-elections. In this paper we propose a Probability Based Adaptive Invoked Weighted Clustering Algorithm (PAIWCA) which can enhance the stability of the clusters by taking battery power of the nodes into considerations for the clustering formation and electing stable cluster-heads using cluster head probability of a node. In this simulation study a comparison was conducted to measure the performance of our algorithm with maximal weighted independent set (MWIS) in terms of the number of clusters formed, the connectivity of the network, dominant set updates, throughput of the overall network and packet delivery ratio. The result shows that our algorithm performs better than existing one and is also tunable to different kinds of network conditions.*

## I. INTRODUCTION

Mobile ad hoc networks (MANETs) consist of mobile devices that form the wireless networks without any fixed infrastructure or centralized administration. In Ad hoc networks, clustering algorithm and select suitable nodes in clusters as cluster heads are so important. This is just because, cluster heads acts as local coordinators and handle various network functions. The clusters are able to store minimum topology information; each CH acts as a temporary base station within its zone or cluster and communicates with other Chs. A clustering scheme should be adaptive to changes with minimum clustering management overhead incurred by changes in the network topology.

The previous research on mobile ad-hoc network has heavily stressed the use of clustering algorithm because clustering simplifies routing and can improve the performance of flexibility and scalability, improved bandwidth utilization, and reduce delays for route strategies. In a clustering structure, the mobile nodes in a network are divided into several virtual zones (clusters).

The process of clustering is never completed without a proper maintenance scheme(Hussein, Salem& Yousef,2008). The objective of cluster maintenance is to preserve as much as of the existing clustering structure as possible. The node movement in the network results in frequent link failure or link establishment between the nodes. This demands cluster member updation to take place from time to time. Moreover, the changing topology and node lifetime / capability (with respect to its available battery power) eliminate the possibility of permanent cluster heads. Thus new cluster heads are required to be elected with the changing scenario. Hence, a well designed clustering algorithm needs to follow a least maintenance overhead phase.

In this paper we propose an Adaptive Invoked Weighted Clustering Algorithm. Which maintains stable clusters. In WCA a node is selected to be the cluster-head with minimum weighted sum of four indices –node degree, distance summation to all its neighboring nodes, mobility and remaining battery power respectively. WCA lacks in knowing the weights of all the nodes before starting the clustering process and in draining CHs rapidly. To solve this problem we propose a probability based adaptive invoked weighted clustering algorithm (PAIWCA).This can enhance the stability of the network by taking battery power of the node into consideration for selecting cluster-heads and for forming clusters. The weight of a node is calculated before the clustering process thus by minimizing the overhead of re-clustering in electing a cluster-head.

## II. RELATED WORKS.

Several original clustering algorithms have been proposed in MANET to choose cluster- heads, namely: (I) Highest-Degree heuristic, (II) Lowest-ID heuristic, (III) Node-Weight heuristic and (IV) weighted clustering algorithm, etc. We will give each of them a brief description as follows.

## A. Highest Degree Algorithm

Highest Degree Algorithm [10], also known as connectivity-based clustering algorithm, was originally proposed by Gerla and Parekh, in which the the degree of a node is computed based on its distance from others. A node x is considered to be a neighbor of another node y if x lies within the transmission range of y. The node with maximum number of neighbors is chosen as a cluster head. The neighbors of a cluster head become members of that cluster and can no longer participate in the election process. Any two nodes in a cluster are at most two-hops away since the cluster head is directly linked to each of its neighbors in the cluster. Basically, each node either becomes a cluster head or remains an ordinary node. Major drawbacks of this algorithm are the degree of a node changes very frequently, the CHs are not likely to play their role as cluster-head for very long time. All these drawbacks occur because this approach does not have any restriction on the upper bound on the number of nodes in a cluster.

## B. Lowest ID Algorithm

Lowest ID Algorithm [14], was originally proposed by each node is assigned a distinct ID .A node, which only hears nodes with ID higher than itself, becomes a Cluster head (CH). The lowest-ID node that a node hears is its cluster head, unless the lowest-ID specifically gives up its role as a cluster head. A node, which can hear two or more cluster heads, is a Gateway. Otherwise the node is an ordinary node. Major drawbacks of this algorithm are its bias towards nodes with smaller ids which may lead to the battery drainage of certain nodes, and it does not attempt to balance the load uniformly across all the nodes.

## C. Node-Weight Algorithm

Node-Weight Algorithm [2],proposed two algorithms, namely distributed clustering algorithm (DCA) and distributed mobility adaptive clustering algorithm (DMAC). In this approach, each node is assigned weights (a real number above zero) based on its suitability of being a cluster head. A node is chosen to be a cluster head if its weight is higher than any of neighbor's weight; otherwise, it joins a neighboring cluster head. The smaller ID node id is chosen in case of a tie. The DCA makes an assumption that the network topology does not change during the execution of the algorithm. To verify the performance of the system, the nodes were assigned weights which varied linearly with their speeds but with negative slope. Since node weights were varied in each simulation cycle, computing the cluster heads becomes very expensive.

## D. Weighted Clustering Algorithm

The Weighted Clustering Algorithm (WCA) was originally proposed by M. Chatterjee et al [6]It takes four factors into consideration and makes the selection of cluster head and maintenance of cluster more reasonable. The four factors are node degree, distance summation to all its neighboring nodes, mobility and remaining battery power respectively. And their corresponding weights are w1 to w4. Besides, it converts the clustering problem into an optimization problem since an objective function is formed.The weight of a node is calculated as

$$w_v = w_1 \Delta_v + w D_v + w M_v + w P_v. \quad (1)$$

Although WCA has proved better performance than all the previous algorithms, it lacks in knowing the weights of all the nodes before starting the clustering process and in draining the CHs rapidly. As a result, the overhead induced by WCA is very high.

## C. Distributed Maximal Weighted Independent Set Algorithm, (MWIS)

Maximal Weighted Independent Set Algorithm [11], was originally proposed by Wandee Wongsason, Chaiyod Pirak and Rudolf Mathar, in which a developed algorithm for deploying the MANETs in realistic large scale networks has been proposed and also MWIS analyzed the complexity of the algorithm under the consideration of MANETs behavior. In MWIS every node $v$ is assigned with $ID_v$ has a certain real-valued weight $w_v > w_u$. This determines its chance to become a clusterhead; the larger weight of a node, the better it is suitable to be a clusterhead. It suggested valid cluster structure,

- every ordinary node is affiliated with exactly one clusterhead as neighbors that has the largest weight among the member node in the independent set represent a clusterhead;
- clusterheads must not be neighbors;
- the maximality guarantees that clusterheads coordinate the entire network.

The dynamic clustering issues have not been discussed in MWIS. It has a drawback of knowing what happens if a new node comes to join a cluster .

## III. PROPOSED WORK

In this section, we present the proposed Probability Based Adaptive Invoked Weighted Clustering Algorithm (PAIWCA).PAIWCA consists of the clustering setup and clustering maintenance phases. Before describing our algorithm in detail, we make the following assumptions:

1. The network topology is static during the execution of the clustering algorithm.
2. Each mobile node joins exactly one cluster- head.

*Phase 1: Clustering setup*

In WCA, the goal is to minimize the value of the sum of all cluster-heads weighted cost. Here a node is selected as cluster head when it minimize a function of four criteria such as degree(number of direct link to its neighbors), sum of distance between cluster head and other nodes, mobility of nodes and battery power of the nodes. When a new node arrives WCA calls the clustering algorithm to determine the weight of the new node for the possibility of being a cluster head. This maximizes overhead in WCA when a new comes. To overcome this drawback of WCA, weight of the node should be known prior to the clustering setup. To achieve this node's weight is calculated using the parameters independent

of the clustering setup. In PAIWCA each node computes its weight value based on the following parameters:

- **Mobility of the node:** Calculate the average speed for every node until the current time T. This gives the measure of the mobility $M_v$.

- **Power consumed:** Determine how much battery power has been consumed at $P_v$. This is assumed to be more for a cluster-head when compared to an ordinary node.

- **Transmission Rate:** Determine the transmission rate for each node at $T_x$. This is assumed to be high for a cluster-head.

- **Transmission Range:** Transmission range for each node ($T_r$) is calculated independently for each node.

- Cluster head probability of node (Chprob).

- The weight $W_v$ for each node is calculated independent of the neighbors and the clusters, using the parameters
    1. Transmission Range, $T_r$
    2. Transmission Rate, $T_x$
    3. mobility of the node, $M_v$
    4. power consumed, $P_v$
    5. C

- **Calculation of weight:** $W_v = (w_1 \times T_r) + (w_2 \times T_x) + (w_3 \times M_v) + (w_4 \times P_v) - chprob$.

- The Values for the constants in our experiments are used as $w_1=0.2, w_2=0.2, w_3=0.05, w_4=0.05$.

- Once the weights of the nodes are calculated before the clustering setup, the node with minimum weight is chosen to be the cluster-head and its neighbors are no longer allowed to participate in the election procedure.

- All the above steps are repeated for remaining nodes which is not yet elected as a cluster-head or assigned to a cluster.

---

*ALGORITHM FOR PROBABILITY BASED ADAPTIVE INVOKED WCA*

$w_1=0.2$;
$w_2=0.2$;
$w_3=0.05$;
$w_4=0.05$;

For each Node
A = Connection between the nodes in a network and the neighbors of the nodes;

$M_v.(X_t,Y_t)$ = average speed for every node until the current time T or the measure of the mobility the node v at instant t;

$P_v$ = The amount of battery power has been consumed. Assumed to be more for a cluster-head when compared to an ordinary node;

$T_r$ = Transmission range of a node;
$T_x$ = Transmission rate of a node;

$W = (w_1 \times T_r) + (w_2 \times T_x) + (w_3 \times M_v) + (w_4 \times P_v)$;

If(W < Minimum)
    Minimum = W;
All the neighbors of the chosen cluster-head are no more allowed to participate in the election procedure;
end for;

---

*Phase 2: Clustering Maintenance*

The second phase is the clustering maintenance. Node movement to the outside of its cluster boundary invokes cluster maintenance. When an ordinary node moves to the outside of its cluster boundary, it is required to find a new CH to affiliate with. If it finds a new CH, it hands over to the new one. If a node moves out of its cluster and does not receive packets from any other node for a specified period of time, it declares itself as CH. If the leaving node is CH, then clustering algorithm is invoked for cluster reorganization.

One disadvantage of WCA is that the clustering algorithm is called on new nodes arrival, which increases the overhead of re-clustering process and the cluster becomes unstable. To avoid this and to increase cluster stability, we propose a probability based approach using **Cluster-Head Probability** of a node.

**ALGORITHM FOR NEWLY ARRIVING NODES**

A= List of connections established with the already existing nodes in the network;
Chprob=node's probability of being a cluster-head;
C-=connection matrix;
If(Chprob(new node)<Chprob(cluster-head))
 Join as a member

If(Chprob(newnode)>Chprob(cluster head))
 Then the new node becomes cluster head of that cluster;

If(In A the new node has no connection with other existing nodes)
 Then new node forms its own Cluster and becomes the cluster-head;

*Cluster-Head Probability:*

Cluster-head probability of a node determines the probability of a node for being a cluster-head. If a newly arrived node has higher probability than the existing cluster-head of a cluster to which it wishes to join, then the newly arrived node becomes the cluster-head for that cluster without disrupting the entire topology which increases the stability of the network. If the newly arrived node has lower probability than the existing cluster-head of the cluster to which it wishes to join then it becomes the member of that cluster-head.
It is calculated as

$$CHprob = Cprob*(Eresidual/Emax) + Tr \qquad (2)$$

1. Eresidual = the estimated current residual energy in the node,
2. Emax is a reference maximum energy (corresponding to a fully charged battery), which is typically identical for all nodes. The CHprob of a node, however is not allowed to fall below a certain threshold Pmin-(e..g.,$10^{-4}$),that is selected to be inversely proportional to Emax.
3. Tr=Transmission range.

*Cluster-Head Probability Threshold*

Battery power of the nodes participating in the Clustering continuously changes. The cluster-head power decreases more rapidly. When the cluster-Head Probability falls below the threshold then the node is no longer be able to peform as a cluster-head and a new cluster-head is chosen.

Check whether the Chprob of a newly arriving node is greater than the existing cluster-head of the cluster to which it wishes to join.

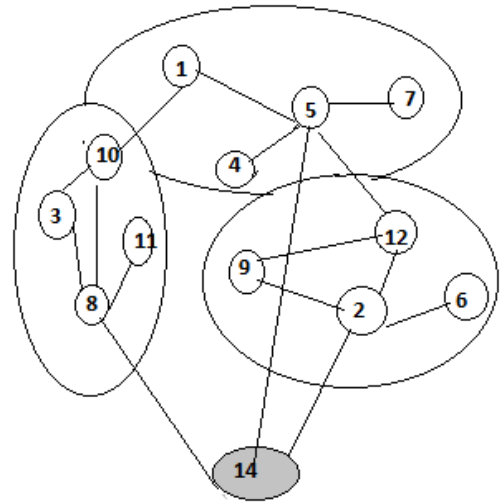

Fig. 1 Arrival of a new node to the existing clusters

In the Fig.1 Node 14 detach from its current cluster and comes into this cluster range. Cluster-Head Probability(**Chprob**) is calculated for node 14 and is compared with the Chprob of the cluster –head to which it is joined, then it is joined to the cluster-head having **Chprob** lesser than the node 14.

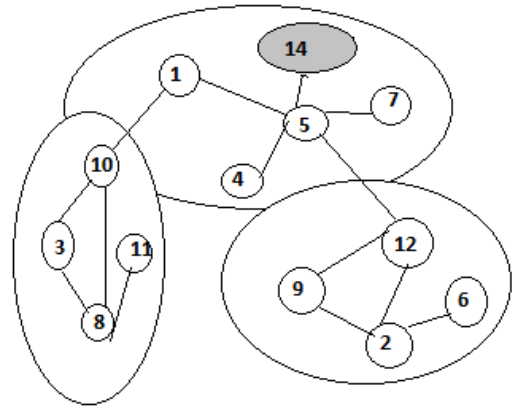

Fig. 2.New node attaches to an existing cluster

The Fig. 2. shows that when the forthcoming node (14)'s Chprob is lesser than the Chprob of cluster head 5, therefore it is allowed to join as a member of the corresponding cluster. Thus overhead of re-clustering is minimized.

---

*ALGORITHM FOR RE-ELECTION OF CLUSTER-HEAD*

A=variable to verify the threshold on the cluster-head's Chprob;
If(A is false)
  Then do not perform Cluster-Head re-election;
Else
  Cluster-Head re-election is called;
All the Member nodes participate in the re-election Procedure using PAIWCA and the node with least weight is selected as the new cluster-head.

---

## IV. EXPERIMENTAL SETUP AND RESULTS

In this section, we present the performance of the proposed PAIWCA, MWIS and WCA by simulation.

The values of the weights used in our simulation are w1=0.2, w2=0.2, w3=0.05, and w4=0.05. The performance of algorithms is measured in terms of the number of clusters formed, dominant set update, the connectivity of the network, throughput of the overall network, packet delivery ratio.

Table 1: Simulation Parameters

| Parameter | Meaning | Value |
|---|---|---|
| N | Number of Nodes | 10 to 300 |
| M X N | Size of Networks | 500 x 500m |
| Max Speed | Maximum Speed | 10 – 100 m/sec |
| $T_R$ | Transmission Range | 5 – 200 m |
| Cluster Member | Number of members for each cluster | 1 – 10 |
| $P_v$ | Initial energy of node | 10-80W |
| $T_r$ | Transmission Range | 10-70m |
| $T_x$ | Transmission Rate | 0.02W |
| M | Movement pattern of nodes | Random way point |
| S | Simultion time | 500 sec |

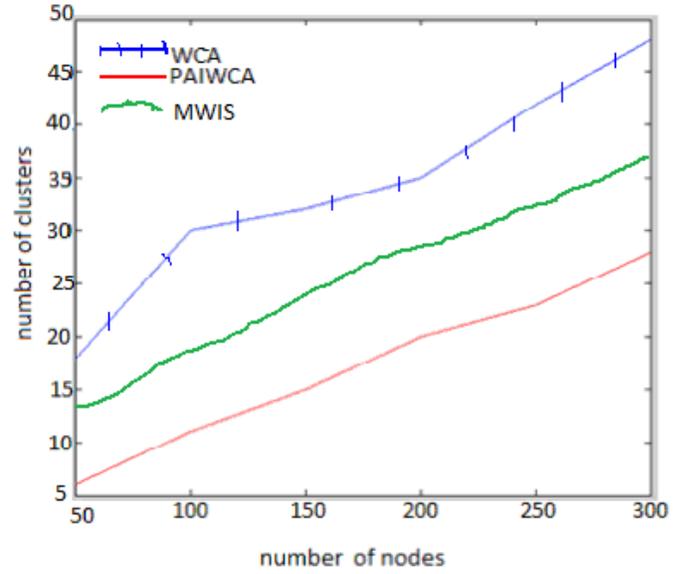

Fig. 3. Comparison between WCA and our proposed work on basis of no of clusters formed

Fig. 3. shows the number of clusters formed for a given number of nodes is plotted. The number of clusters formed should be reduced. In PAIWCA the number of clusters formed is comparatively less. Minimum number

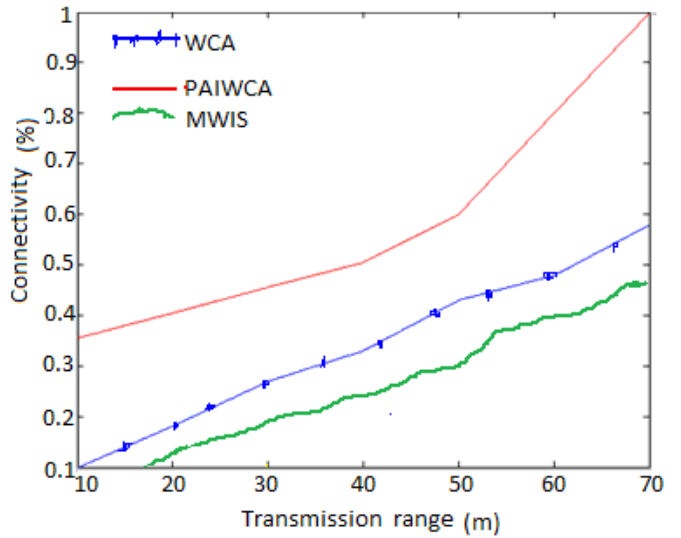

Fig. 4. Comparison between WCA and our proposed work on the basis of Transmission range vs Network connectivity, N=50.

Figure 4 shows the connectivity of the network for WCA and PAIWCA. The connectivity is defined as the probability that a node is reachable from any other node. For a single component graph, any node is reachable from any other node and the connectivity is 1. If the network does not results in single component graph, the all the nodes in the largest component can communicate with each other. Thus,

$$\text{Connectivity} = \frac{\text{size of largest component}}{N}$$

The cluster-head power should be high to yield a connected graph. In PAIWCA cluster-head is chosen in such a way to have maximum battery power using Cluster-head probability to yield a connected graph within a shorter transmission range.

The dominant set is the set of cluster-heads. In WCA the dominant set update is high when compared to PAIWCA. In PAIWCA the cluster-head change rate can be minimized by minimizing the cluster-head re-election process using the Chprob of newly arrived node to determine whether to call the cluster-head election process or not. As the clusterhead can be used as a repository for the knowledge of the cluster and a coordinator of the cluster operations ,it should posses maximum battery  The node with maximum value of Chprob will have maximum battery power.Thus stability of the network is increased by reducing dominant set update.

Figure 6 represents the throughput of forwading message in PAIWCA .In PAIWCA  throughput of forwading  message is increasing  and it is stable.

Probability Based Adaptive Invoked Weighted Clustering Algorithm (PAIWCA) shows 100% connectivity as the transmitting range increases.Mobitlity analysis in NS2 simulated environment with 100 nodes is given with different Pause time values which influences  the node mobility in the random waypoint model.The lower the Pause time higher the mobility.The number of sources is 100 with varying Pause time is used forn this analysis.The Packet  delivery ratio is based on the number of packets delered and number of packets sent. Higher the Packet Delivery Ratio (PDR) better is the performance of the cluster.

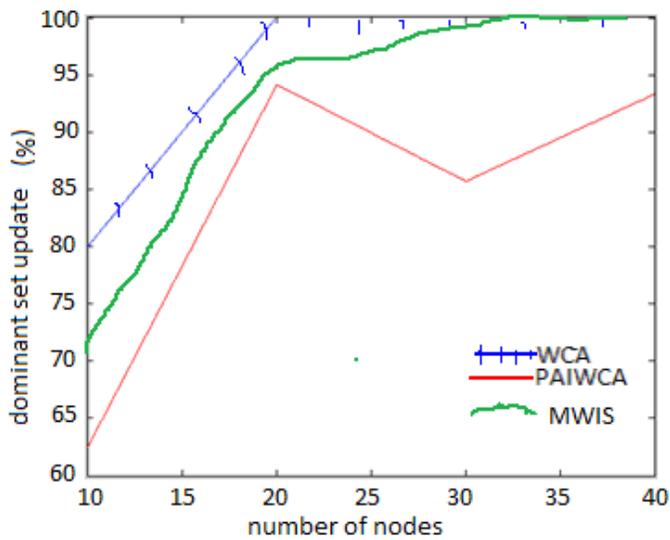

Fig. 5.  Dominant set update vs number of nodes.

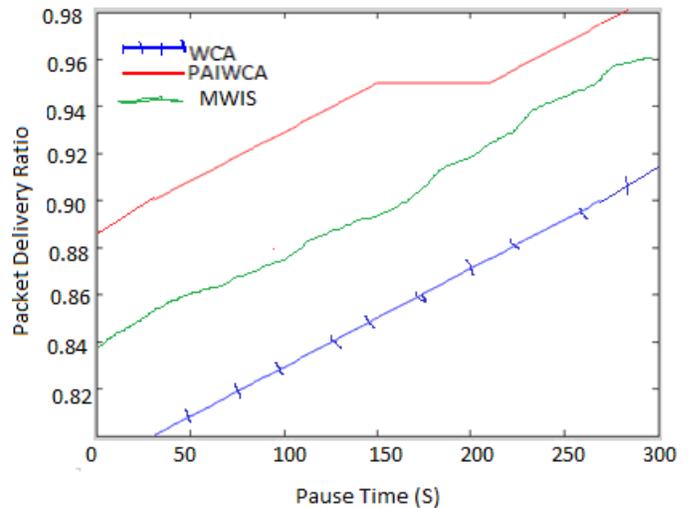

Fig. 7.  Packet Delivery Ratio  vs Pause time

Only under the dynamic condition the packets collide or the packets are lostFor WCA and Lowest ID algorithm even as the nodes become static all the packets are not delivered.

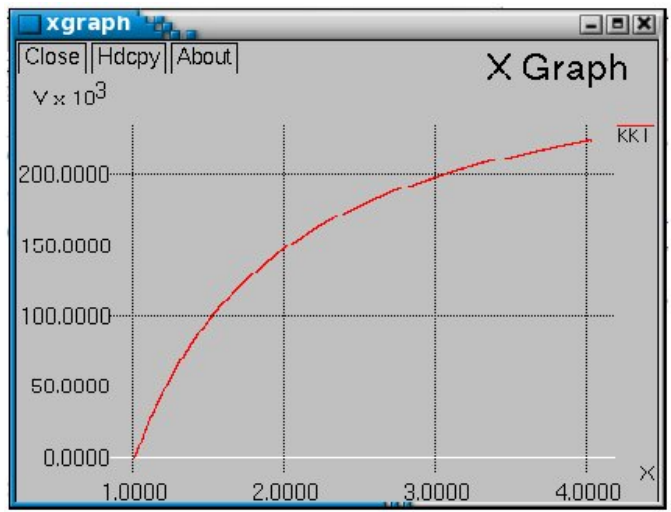

Fig. 6.  This graph  represents  PAIWCA output.Throughput vs simulation time

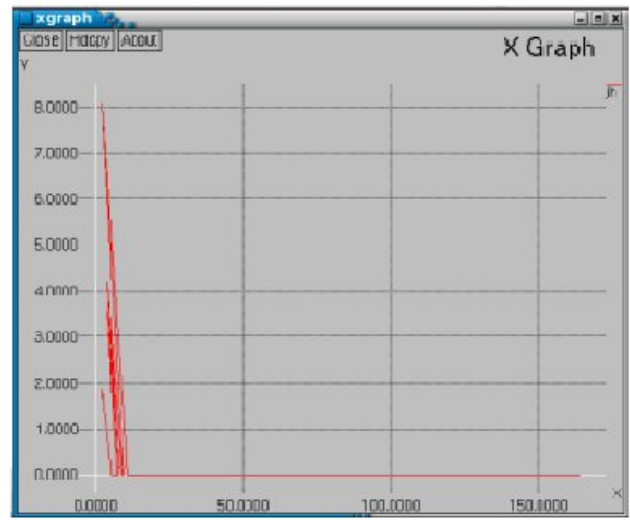

Fig. 7.  This graph  represents  PAIWCA output.Throughput vs simulation time

The communication delay of a node in MANET mainly includes queue delay, process delay, and transmission delay and so on. For a single data packet, the process delay and transmission delay are relatively fixed, while the queue delay is subject to the network congestion condition which has a wide range. To compute average end-to-end delay, we consider only delivered packets because this delay can be considered as infinite for lost packets. The end-to-end delay increases when the simulation time increases.

V. CONCLUSION

In this paper we have presented a Probability based Adaptive Invoked Weighted Clustering Algorithm(PAIWCA) that can be applied in MANET to improve upon their stability, to reduce the dominant set update, to increase the connectivity of the network, to increase the throughput and packet delivery ratio. WCA has the major disadvantage of knowing weights of the nodes before forming the clusters thereby increasing the overhead of re-election when new nodes arrive or when the cluster-head moves out. PAIWCA calculates weights of the nodes before forming the clusters thereby minimizing the overhead of re-election process. The performance of the proposed PAIWCA demonstrated that it outperforms WCA in terms of connectivity and stability.